\documentstyle[12pt,epsfig]{article}

\newcommand{\beq}{\begin{equation}}
\newcommand{\eeq}{\end{equation}}
\newcommand{\beqa}{\begin{eqnarray}}
\newcommand{\eeqa}{\end{eqnarray}}

\newcommand{\labell}[1]{\label{#1}}
\newcommand{\reef}[1]{(\ref{#1})}
\newcommand{\lsim}{\mathrel{\raisebox{-.6ex}{$\stackrel{\textstyle<}{\sim}$}}}
\newcommand{\gsim}{\mathrel{\raisebox{-.6ex}{$\stackrel{\textstyle>}{\sim}$}}}

\def\ie{{\it i.e.,}\ }
\def\eg{{\it e.g.,}\ }

\let\bar=\overbar

\def\etal{{\it et al.}}
\def\ie{{\it i.e.}}
\def\eg{{\it e.g.}}

\def\Dslash{\not{\hbox{\kern-4pt $D$}}}
\def\dslash{\not{\hbox{\kern-2pt $\del$}}}

\def\msb{{\bar{\ssstyle M \kern -1pt S}}}

\begin{document}

\thispagestyle{empty}
\rightline{\hfill FERMILAB-Pub-00/014-T}
\vskip -.1ex
\rightline{\hfill McGill/00-11}
\vspace*{2cm}

\begin{center}
{\bf \LARGE Gravity in a Box}
\vspace*{1cm}

Joseph Lykken$^{a,}$\footnote{E-mail: lykken@fnal.gov}, 
Robert C. Myers$^{b,}$\footnote{E-mail: rcm@hep.physics.mcgill.ca}
and Jing Wang$^{a,}$\footnote{E-mail: jingw@fnal.gov}
\vspace*{0.3cm}

$^a${\it Theory Department, Fermi National Accelerator Laboratory}\\
{\it P.O. Box 500, Batavia, IL 60510, USA}\\
\vspace*{0.2cm}
$^b${\it Department of Physics, McGill University}\\
{\it 3600 University Avenue, Montr\'eal, QC, H3A 2T8, Canada}\\
\vspace{2cm}
ABSTRACT
\end{center}
We consider a brane-world construction which incorporates
a finite region of flat space,
``the box,'' surrounded by a region of anti-de Sitter space.
This hybrid construction provides a framework which interpolates
between the scenario proposed by Arkani-Hamed, Dimopoulos and
Dvali, and that proposed by Randall and Sundrum. 
Within this composite framework, we investigate the effects
of resonant modes on four-dimensional gravity.
We also show that, on a probe brane in the anti-de Sitter region,
there is enhanced production of on-shell nonresonant modes.
We compare our model to some recent attempts to
incorporate the Randall-Sundrum scenario into superstring theory.

\vfill
\setcounter{page}{0}
\setcounter{footnote}{0}
\newpage

\section{Introduction}

There has been a renewed interest by the high energy community in
the possibility
that spacetime may have a dimension larger than four. 
The essential new concept which sparked this interest is the notion that
only gravitational
excitations propagate through the full spacetime, while all
of the observed Standard Model particles are confined to a three-brane.
Such a scenario may be motivated by ideas naturally arising in string
theory\cite{stringa,stringb}. One of the most exciting phenomenological
implications
of these brane-world scenarios is that the fundamental scale of gravity
may be reduced from the four-dimensional Planck scale of $10^{16}$ TeV
to as little as 1 TeV.

The wave of research exploring these scenarios has moved
forward on two seemingly disjoint fronts.
In the original work of Arkani-Hamed \etal\cite{Arkani,Anton},
there are large extra dimensions, which may be as big as a fraction
of a millimeter.
For simplicity, the extra compact dimensions are usually assumed to be
toroidal,\footnote{Compactifications using hyperbolic manifolds
were considered in ref.~\cite{stark}.}
and then the gravitational fluctuations have a simple Fourier mode expansion.
The zero-mode in this expansion is interpreted as the massless
four-dimensional
graviton responsible for the observed long-range effects of gravity.

More recently, Randall and Sundrum\cite{small,big} proposed a scheme in which
a five-dimensional spacetime contains strongly gravitating three-branes which
then produce a warped or nonfactorizable geometry. With a particular
tuning of the bulk (negative)
cosmological constant and the brane tension, the induced geometry on
the three-branes is just flat four-dimensional Minkowski
space.\footnote{One can also induce cosmological geometries on the
three-branes by varying these parameters\cite{cosmo}.}
There is also a massless four-dimensional graviton, but its wave function is
{\it localized} in the nonfactorizable geometry of the extra dimension. 
In this construction, the size of the extra dimension is unconstrained:
it could
be either very small\cite{small} or very large\cite{big}.
This scenario can also be extended to
cases with more than one extra dimension\cite{more}.

In this paper, we study a brane setup
which interpolates between these two classes of scenarios,
with a single extra dimension. Our five-dimensional construction
contains a nonfactorizable geometry which is
asymptotically anti-de Sitter (AdS), as in the Randall-Sundrum (RS) scenario.
However, as in the framework of Arkani-Hamed--Dimopoulos--Dvali 
(ADD), there is a completely flat region of finite width
-- ``the box'' -- bounded by three-branes.\footnote{An interesting
construction based on the opposite hybrid picture,
where a finite AdS region is surrounded by an asymptotically flat
space, was recently proposed\cite{Gregory}. Whether this yields a viable 
effective theory of four-dimensional gravity is currently under
scrutiny\cite{debate}.} By varying the relative scales
of the flat box
and the AdS regions, this setup interpolates between a limit
in which it reproduces the RS construction and another where it yields
the ADD scenario. We will study the low-energy effective theory
of four-dimensional gravity that arises on the branes. In particular,
we will focus on the effects of ``resonant modes,'' which are metric
modes which have enhanced support inside the box region.

\section{Background Geometry}

We begin by considering a five-dimensional nonfactorizable background geometry
whose metric takes the following form in Poincar\'e coordinates: 
\beq
\labell{metric}
ds^2=e^{-2A(y)}\eta_{\mu \nu}dx^{\mu}dx^{\nu} + dy^2 \;. 
\eeq
For simplicity, we will restrict our attention to geometries
(both the background
and the metric fluctuations in the following section) which are
reflection symmetric around $y=0$. Thus we are essentially
considering a $Z_2$ orbifold (which may be either compact or noncompact). 
We will also be interested in the case where the geometry is
asymptotically AdS, \ie\ for large $|y|$,
\beq
A(y) \rightarrow k|y| \;,
\eeq
where $k$ is the inverse of the AdS radius of curvature.

The simplest example of a brane setup which produces a background geometry
of this type is to have a number of three-branes with
positive tension superimposed at $y=0$. We refer to these branes
collectively as the ``Planck brane,'' and we will designate this simple setup
as the {\it RS limit} --- see below.
We consider the five-dimensional gravity action  
\beqa
\labell{action}
S & = & S_{bulk} + S_{brane} \;, \\
{\rm with}\qquad
S_{bulk} & = & \int d^4x \,dy\, \sqrt{-g} \,( 2M_5^3 R - \Lambda) \;,
\nonumber\\
S_{brane} & = & - \int d^4x\, \sqrt{-g_{4}}\,V_{P} \;,
\nonumber
\eeqa
which should be considered the leading terms in a low-energy effective
action. Here, $g_{MN}~(M,N=0,\ldots 4$) is the five-dimensional metric, while
$(g_{4})_{\mu\nu}$ $(\mu,\nu=0,\ldots 3$) is the induced
metric on the Planck brane. Also, $M_5$, $\Lambda$
and $V_P$ denote the five-dimensional Planck scale, the (negative) bulk
cosmological constant and the (total) brane tension, respectively.
Finding a solution which is Poincar\'e invariant in four dimensions
requires that the tension $V_P$ is tuned relative to the cosmological
constant $\Lambda$. That is, we set $V_{P} = 
\sqrt{24 M_5^3 |\Lambda|}$, as in \cite{small}.
The solution of the five-dimensional Einstein equations in this RS setup is
then given by the metric (\ref{metric}), with $A(y)~=~k|y|$, where
\beq
\labell{kvalue}
k^2=-\frac{\Lambda}{24M_5^3} \;.
\eeq 
This background geometry is simply two AdS regions glued together
along the surface $y=0$ with the Planck brane supporting the appropriate
discontinuity in the extrinsic curvature across the gluing surface.

Now imagine splitting
the Planck brane into two sets and pulling them away symmetrically
from $y=0$ to $y=\pm y_0$. Given a symmetric division of the Planck brane,
the tension of each of the two subsets is $V_P/2$. 
We also require that the
region of the bulk space between the branes (\ie\ with $|y|<y_0$)
is in a new vacuum where the bulk cosmological constant vanishes.
Then the three-branes at $y=y_0$ remain flat with the same tuning
of the tension $V_P$ given above.
Given the vanishing of $\Lambda$ in the small $y$ region, solving Einstein's
equations in this part of the spacetime will yield
a slice of flat five-dimensional Minkowski space.
The full solution becomes the metric (\ref{metric}) with
\beq
\labell{classicalsol}
A(y) = \frac{1}{2} k |y-y_0| + \frac{1}{2}k|y+y_0| - ky_0 \;, 
\eeq
where $k$ has the same value as in eqn.~(\ref{kvalue}).
The resulting picture is then a flat ``box'' glued between two
AdS regions. Of course, the
RS limit is now that in which the box shrinks to zero size,
\ie\ $y_0\rightarrow0.$    

\section{Graviton Modes} 

When linearized metric fluctuations are included, the geometry
takes the form
\beq
\labell{fluctu}
ds^2= (e^{-2A(y)}\eta_{\mu \nu}+ h_{\mu \nu})dx^{\mu}dx^{\nu} + dy^2 \;.
\eeq
In the following, we will work in a gauge where $\partial^{\mu}h_{\mu \nu}
=h^{\mu}_{\mu}=0$. We will not consider the metric fluctuations $h_{55}$ and
$h_{5\mu}$ (which are pure gauge for the case where $y$ has an infinite
range).
It is useful to define a conformal coordinate $z$ by
$z \equiv sgn(y) [(e^{k(|y|-y_0)}-1)/k+y_0]$
when $|y| \ge y_0$, and $z\equiv y$ when $|y| \le y_0$.
Given our background solution \reef{classicalsol}, the geometry \reef{fluctu}
then becomes
\beq
\labell{newnet}
ds^2 = \left\lbrace
\matrix{(\eta_{\mu \nu}
+ h_{\mu \nu})dx^{\mu}dx^{\nu}+dz^2&{\rm for}&|z|\le z_0\cr
        {1\over(k\tilde{z})^2}
           \left[\,(\eta_{\mu \nu}+ \hat{h}_{\mu \nu})dx^{\mu}dx^{\nu}
            +dz^2\,\right]
              &{\rm for}&|z|\ge z_0 \cr}\right.
\eeq              
where $z_0=y_0$, $\tilde{z}~=~|z|-z_0+1/k$, and $\hat{h}_{\mu \nu}
= e^{2A(z)}h_{\mu \nu}$.

Now solve the linearized Einstein equations for $h_{\mu\nu}$ with
separation of variables using an ans\"atz of the form:
$h_{\mu \nu}~=~e^{ip \cdot x} e^{-A(z)/2} \psi_m(z) \epsilon_{\mu \nu}$.
Here $\epsilon_{\mu \nu}$ is a constant polarization tensor. The
four-dimensional profile of these solutions is a plane wave with an effective
four-dimensional mass: $m^2=-p^2$. Solving the linearized equations
is now reduced to a one-dimensional
Schr\"odinger problem:
\beq 
\labell{waveequ}
\left[ -\frac{1}{2} \partial_{z}^2 + V(z) \right] \psi_m(z) 
= \frac{1}{2}m^2 \psi_m(z) \;, 
\eeq
where the  potential $V(z)$ is given by
\beq
\labell{V(z)}
V(z) = \frac{15 k^2}{8(k|z|-kz_0+1)^2} \theta(|z|-z_0) - 
\frac{3k}{4}\delta(|z|-z_0) \;. 
\eeq
With these definitions, the natural norm\footnote{This norm is inherited
from the standard relativistic or ``Klein-Gordon'' inner product of
the five-dimensional graviton fluctuations in the given background
geometry \reef{metric}.} for the
profile in the fifth dimension is simply $\int\! dz\, |\psi_m(z)|^2~=~1$.  

The solution to eqn.~(\ref{waveequ}) is a combination of plane waves in
the box and Bessel functions in the AdS region: 
\beq
\psi_m(z)=\left\lbrace
\matrix{B_m\,\cos mz&{\rm for}&|z|< z_0 \;,\cr
        N_m (k\tilde{z})^{1/2}\left[ Y_2(m\tilde{z})+
        L_m\, J_2(m\tilde{z})\right]
              &{\rm for}&|z|> z_0 \;,\cr}
\right.
\labell{psim}
\eeq
where $\tilde{z}$ is as defined below eqn.~\reef{newnet},
while $B_m$, $N_m$, and $L_m$ are $m$-dependent coefficients.
$L_m$ is determined by the jump condition at the Planck brane, 
\beq
\labell{lfun}
L_m=-{Y_1(m/k)+Y_2(m/k)\,\tan(mz_0)\over J_1(m/k)+J_2(m/k)\,\tan(mz_0)} \;,
\eeq
while $B_m$ is fixed by requiring the $\psi_m$ to be continuous at $z_0$,
\beq
\labell{contfun}
B_m\cos(mz_0)=
N_m\left[Y_2(m/k)+L_m\,J_2(m/k)\right] \;.
\eeq
This leaves $N_m$ to be fixed by imposing the appropriate normalization
of the profile.

We can introduce a second boundary, \ie, a second $Z_2$ orbifold surface,
at some finite $z=z_c$ in the AdS region, by inserting branes of
negative tension
$V_{N} = -V_{P}$, as in \cite{small}. In this case, the
mass spectrum becomes discrete because a second jump condition must
be imposed on the mode functions at the negative tension branes.
The latter provides an independent equation fixing $L_m$, which
combined with eqn.~\reef{lfun} yields
\beq
\labell{quantization}
 \frac{Y_1(m/k) + Y_2(m/k)\tan(mz_0)}{J_1(m/k) + J_2(m/k)\tan(mz_0)} 
= \frac{Y_1(m\Delta z+m/k)}{J_1(m\Delta z+m/k)},
\eeq 
where $\Delta z=z_c - z_0$. 

If we take $\Delta z\rightarrow 0$, the two sets of branes
``annihilate'' leaving behind a braneless compactification
on $S^1/Z_2$. This is a smooth limit in our low-energy description.
We will refer to this as the {\it ADD limit}. For this limit to be 
phenomenologically
viable, one would have to assume that there are Standard Model fields
living on the orbifold surfaces, in analogy to the M-theory
scenario proposed in \cite{stringa}. Alternatively, 
one could introduce a probe brane (which does not disturb the
background geometry) to support the Standard Model fields.

In all of the cases that we are considering, there is
a normalizable zero-mode:
\beq
\labell{psinot}
\psi_0(z)=\left\lbrace
\matrix{B_0&{\rm for}&|z|< z_0 \;,\cr
        B_0\left(k\tilde{z}\right)^{-3/2}
            &{\rm for}&|z|> z_0 \;.\cr}
\right.
\eeq
The normalization condition determines $B_0$: 
\beq
\labell{normnot}
B_0=\left({k\over 2kz_0+1-e^{-2k\Delta z}}\right)^{1/2} \;.
\eeq
where the last term in the denominator vanishes in the case
of an infinite fifth dimension, \ie\ when $\Delta z\rightarrow\infty$.

The existence of this zero-mode is also evident as follows: One finds
that the five-dimensional metric \reef{metric} remains a solution of the
field equations derived from eqn.~\reef{action} when the flat metric
$\eta_{\mu\nu}$ is replaced by a general Ricci-flat metric
$\tilde{g}_{\mu\nu}(x)$.
That is, the five-dimensional equations of motion are still satisfied
as long as the brane metric satisfies the four-dimensional Einstein
equations $R_{\mu\nu}(\tilde{g})=0$. The zero-mode solutions appearing
in the linearized
calculations above are the usual gravity waves appearing in a perturbative
analysis of these four-dimensional gravity equations.

Using this general nonlinear ans\"atz, we can also calculate the effective
four-dimensional Planck scale for observers on the three-branes
at $z=z_0$. We simply
insert our ans\"atz, eqns.~\reef{metric} and \reef{classicalsol}, with
$\tilde{g}_{\mu\nu}(x)$ replacing $\eta_{\mu\nu}$, into the five-dimensional
action \reef{action}. Now integrating over $y$ leaves an effective
four-dimensional Einstein action with an overall coefficient of
$2M_{\rm Planck}^2$ where
\beq
\labell{plank}
M_{\rm Planck}^2={M_5^3\over B_0^2}={M_5^3\over k}
\left(2kz_0+1-e^{-2k\Delta z}\right)\ .
\eeq
Again, the final term in the second factor vanishes for the case of
an infinite fifth dimension.

\section{An Infinite Fifth Dimension}

First we consider the case of the box embedded in
an AdS space without boundary. There is a continuum spectrum of
massive gravity modes. Unlike the zero-mode \reef{psinot}, a unit norm
can not be imposed on these modes because the $\psi_m(z)$ have
plane wave behavior asymptotically. Instead these modes are
given $\delta$-function
normalization, \ie\ $\int \psi_m^*(z)\,\psi_{m'}(z)\,dz =\delta(m-m')$.
Comparing to eqn.~\reef{psim}, this implies:
\beq
\labell{mans}
N_m^2={1\over2}{m\over k}{1\over 1+L_m^2} \;.
\eeq
The coefficient $B_m$ is obtained from eqn.~(\ref{contfun}).

Now the essential question we would like to answer is the extent to
which gravity on the branes at $z=z_0$ is four-dimensional. In particular,
we will examine how the gravitational potential is modified by the
massive gravity modes in the bulk. Following \cite{big},\footnote{See
\cite{pot} for a more extensive discussion.} the gravitational potential
between two test masses, $m_1$ and $m_2$, separated by a distance $r$ on the
Planck brane, takes the form 
\beq
\labell{newt}
U(r)=-{G_4m_1m_2\over r}\left(1+\int dm\, \rho(m)
\, e^{-mr} \right) \ .
\eeq
Here the four-dimensional Newton's constant is defined as
\beq
G_4 \equiv {1\over32\pi M_{\rm Planck}^2}=
{1\over32\pi M_5^3} {k\over 2kz_0 + 1}\;\; .
\labell{ournewt}
\eeq
We have also introduced a relative density of states
\beq
\labell{dense}
\rho(m) \equiv {|\psi_m(z_0)|^2\over |\psi_0(z_0)|^2 }
\eeq
for the massive modes. Now for large distances, the dominant
contributions to the integral over the massive modes will come
from $m<1/r$. In general, we only consider distances
$r>1/k$ where these dominant contributions come from $m<k$.
This latter restriction is made because from the form of the
potential \reef{V(z)} in the Schr\"odinger equation \reef{waveequ},
it is clear that modes with $m\gsim k$ will not be suppressed at
the Planck brane. The strong coupling of these bulk modes
indicates that we should expect that the approximately four-dimensional
character of gravity on the brane must break down for $r<1/k$.

Now if the size of the box is small or comparable to the AdS scale,
\ie, $kz_0\lsim 1$, it is not hard to show that the leading corrections
to the long-range gravitational potential are in fact identical
to those in the RS limit. One finds
\beq
\labell{densa}
\rho(m)\simeq {1\over 2}{m\over k^2}
\eeq
and so is independent of $z_0$. The final result for the gravitational
potential is
\beq
\labell{newtan}
U(r)\simeq-{G_4m_1m_2\over r}\left(1+{1\over 2k^2r^2} \right) \ .
\eeq
Hence there is a power law correction to the four-dimensional
Newtonian potential, which is controlled by the AdS scale $k$.

\begin{figure}
\centerline{
\hbox{
\epsfxsize=5.0truein           
\epsfbox[70 32 545 740]{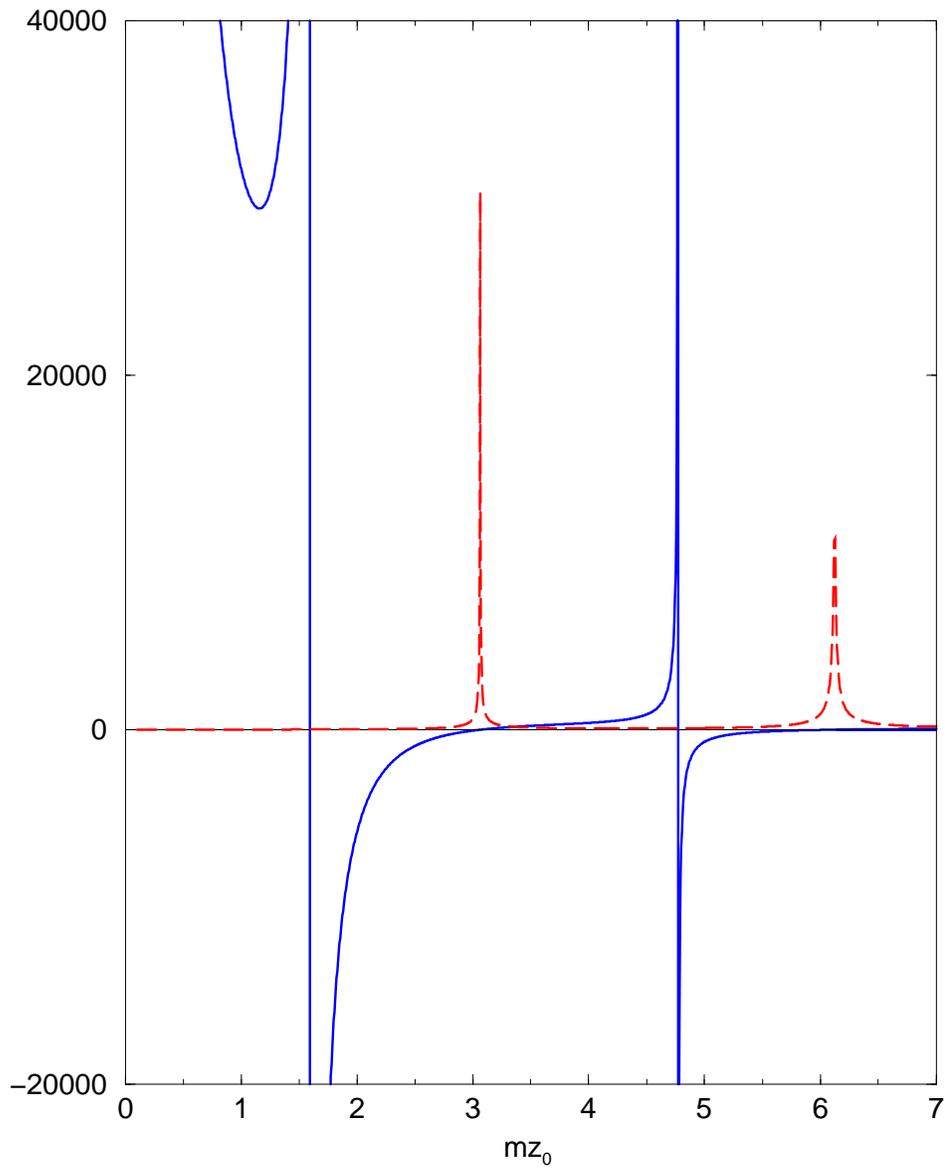}
}
}
\caption{$L_m$ (solid line) and $|B_m|$ (dashed line)
as functions of $mz_0$. $kz_0=20$ is chosen for the plot,
and $B_m$ is rescaled by a factor of 2000.
}  
\end{figure}

In the regime of a large box, \ie, $kz_0\gg 1$, more interesting
behavior is found. In Fig.~1,
we plot $L_m$ and $|B_m|$ as a function of $mz_0$ for fixed $z_0$.
The figure illustrates the generic behavior in this regime. That is,
$B_m$ goes through periodic extrema as $m$ increases,
while at these $m$ values
$L_m$ is very close to $0$. These modes at the extrema of $B_m$,
which have enhanced support
inside the box, are identified as the {\it resonant
modes}. Numerically, we find that these resonances
persist for relatively small boxes, \eg, $kz_0\simeq10$, but to produce
precise analytic results below, we will restrict our attention to
the regime $kz_0\gg 1$.

In our long-range or low-energy approximation,
we restrict our attention to light modes:
$m/k \ll 1$. The condition $L_m\sim 0$ for
resonant modes then reduces to
\beq
\labell{small}
\tan mz_0\simeq -{m\over2k} \; ,
\eeq
which means that $mz_0\simeq n\pi$
for some positive integer $n$. Near the zero, the tangent function
is essentially linear, so to leading order we can write
\beq
\labell{linear}
\tan mz_0\simeq mz_0-n\pi
\eeq
and we find
\beq
\labell{approxL}
L_m\simeq {4\over\pi}{k^3\over m^3}(
x - 2\pi n) \;,
\eeq
where $x \equiv (2kz_0+1)(m/k)$. Thus the resonance mass is
\beq
\labell{resm}
m_n\simeq{2\pi nk\over2kz_0+1}\simeq {\pi n\over z_0} \; .
\eeq
Now near the resonant masses, the value of the wave
functions on the Planck brane can be written as
\beq
|\psi_m(z_0)|^2\simeq|B_m|^2\simeq |N_m Y_2(m/k)|^2
\simeq {2\over\pi}{Q\over 1+Q^2 (x-2\pi n)^2} \; ,
\labell{peak}
\eeq
where
\beqa
Q&\equiv&{4\over\pi}\left({k\over m}\right)^3 \; .
\labell{defxq}
\eeqa
Note that the extremal value of $|\psi_m(z_0)|^2$ is proportional
to $Q$, while the width at half-maximum is $\Delta x=2Q$. Hence from
eqn.~\reef{defxq}, one sees that the peaks in eqn.~\reef{peak} become
higher and narrower for smaller resonant masses
--- a feature which can be observed in the plot of
$|B_m|$ in Fig.~1.
With $m/k\ll 1$, $Q$ is large and the expression in
eqn.~(\ref{peak}) is a good approximation to 2 times
a delta function.
Hence the correction to the Newtonian potential \reef{newt} becomes
\beqa
\int dm\, \rho (m) \, e^{-mr}
&\simeq&{2\over\pi}\sum _{n=1}^{\infty}
\int dx {Q\over 1+Q^2x^2} e^{-m_nr}
\nonumber\\
&=&2\,\sum _{n=1}^{\infty}
e^{-m_nr} \; ,
\labell{final}
\eeqa
where we have treated both $Q$ and $m$ in the exponential as
slowly varying functions. We have also used
 $dm = |\psi_0(z_0)|^2dx$. The sum over $n$ is actually cut off
at $n \sim kz_0$. 

Within our approximations then, the continuum of states around
a resonant mass makes a contribution to the Newtonian potential as though
there were a single {\it discrete} normalizable mode with mass
$m=m_n$. The total gravitational potential on the Planck brane becomes 
\beq
\labell{totalcorr}
U(r) \simeq -\frac{G_4 m_1m_2}{r}\left(1 + \frac{1}{2k^2r^2} 
+ 2 \frac{e^{-r/r_e}}{1-e^{-r/r_e}}\right), 
\eeq
where the second term comes from the 
nonresonant continuum modes, and the third
term comes from summing over the contributions at all
of the resonant masses. The effective length
scale appearing in this last term is: $r_e \simeq z_0/\pi$.
Therefore within the large box regime with $kz_0>>1$, these
resonant mode contributions are in fact the leading contributions
to the Newtonian gravitational potential.

\section{A Finite $Z_2$ Orbifold}

In the case where a second orbifold surface is introduced at $z=z_c$, the
spectrum of the gravity modes becomes discrete as determined by 
eqn.~(\ref{quantization}). Hence the details of the spectrum are controlled
by the three different scales,
$z_0$, $\Delta z$, and $1/k$, entering this quantization constraint.

First consider the regime: $\Delta z \gg z_0,1/k$. In this case
the spacing of the masses is very small, $\delta m \simeq \pi/\Delta z$,
which can be seen as follows: on the right-hand-side (r-h-s)
of eqn.~\reef{quantization},
we can approximate the Bessel functions with their asymptotic plane
wave forms to give
\beq
\labell{rhs}
 \frac{Y_1(m/k) + Y_2(m/k)\tan(mz_0)}{J_1(m/k) + J_2(m/k)\tan(mz_0)} 
\simeq \tan(m\Delta z+m/k-3\pi/4)\ .
\eeq 
Now the r-h-s is a rapidly varying function of $m$ compared to the l-h-s.
In particular, the r-h-s varies from $-\infty$ to $\infty$ as $m$ increases
by $\pi/\Delta z$. Hence this constraint \reef{rhs} will be satisfied
once in every interval $n\pi/\Delta z<m<(n+1)\pi/\Delta z$. With such a
tight spacing
of the mass spectrum, the physics is still essentially unchanged from the
case with an infinite fifth dimension discussed in the previous section.
In particular, if we are also in the regime where $z_0 \gg 1/k$, the
modes in the discrete spectrum satisfying $m\simeq n\pi/z_0$ will have
enhanced support in the box region. These resonant modes will then
dominate the corrections to the four-dimensional gravitational potential.

Let's now consider the situation in which the size of the box is much
larger than the size of the AdS space,
\ie\ $z_0 \gg \Delta z \gg 1/k$.
For modes with $m \ll k$, eqn.~(\ref{quantization})
gives:
\beq
\labell{s1m}
\tan(mz_0) \simeq - \frac{m}{2k} \ll 1, 
\eeq
thus $m \sim n \pi/z_0$ and $\cos(mz_0) \sim 1$. Note that this
result is precisely the condition for the resonances \reef{small}
in the case of the infinite fifth dimension.

With $m\tilde{z} \ll 1$, $Y_{2}$ dominates the shape of the wave
function outside the box. The normalization condition can be written
$I_1 + I_2 =1$, where $I_1$ is the contribution from outside
of the box:
\beq
\labell{s1i1} 
I_1 \simeq  k N_m^2 \int_{z_0}^{z_0 + \Delta z} dz\, \tilde{z}\,
|Y_2(m\tilde{z})|^2
\simeq
N_m^2\, \frac{16}{\,\pi^2}\,\frac{k^4}{m^4}\,\frac{1}{k}\;,
\eeq
and $I_2$ is the contribution from inside the box:
\beq
\labell{s1i2} 
\begin{array}{ccl}
I_2 &\simeq & B_m^2 z_0 \;, \\
 &\simeq& N_m^2\, \frac{16}{\,\pi^2}\,\frac{k^4}{m^4}\,z_0\;.
\end{array}
\eeq
Since we are considering $z_0 \gg 1/k$, $I_2 \gg I_1$
and $B_m \simeq 1/\sqrt{z_0}$.  

Therefore, $\rho(m) \simeq 1/2$ on the Planck brane. Thus up to a factor
of order one, the leading correction to the gravity potential is the same
as the third term in eqn.~(\ref{totalcorr}), with the same effective
length scale (given the matching between eqns.~\reef{small} and \reef{s1m}).
Hence in this regime, the large box again mimics the situation with one extra
flat dimension of size $z_0$.  

\section{No Unusually Large Boxes}
In the case just described the size $z_0$ of the box cannot be
larger than a millimeter without conflicting with Cavendish type
experiments which directly measure the Newtonian potential.
Now we pause to inquire whether it is possible to weaken this
limit, by somehow introducing a large wave function suppression
for the resonant modes. 

One obvious strategy is to change the location of our observer
in the fifth dimension; to this end we could confine the Standard
Model to a probe brane, as in the scenario of
Lykken and Randall\cite{Lykken}. However locating the probe
brane in the interior of the box is no help, and locating it
in the interior of the AdS region is no better,
since for light modes, $Y_2$ dominates the wave function and simply tracks
the zero-mode. Thus
$\rho(m) \sim 1$ still applies.

The difference between the modes in eqn.~(\ref{s1m}) and that of RS is that
in the first case the behavior of these resonant
modes is dominated by function $Y_2$
in both the small {\it and} large $z$ regions
of AdS, while in the RS case,
$\psi_m$ is dominated by $Y_2$ when it is close to the Planck brane, but by
$J_2$ when it is far away.
The suppression of the massive graviton wave functions
on the Planck brane in the RS limit
comes as a balance between the normalization factor, determined
mainly by $J_2$ term, and the behavior of $Y_2$
at the Planck brane.

However, one could imagine another situation where
$\psi_m$ is dominated by $J_2$ thoughout the AdS region.
This would cause
suppression of the
massive graviton wave function on the Planck brane since
$J_2(m/k) \sim (m/k)^2$ which is small for the light modes.
We have found that there indeed
exist such modes if one assumes that $\Delta z \sim z_0 \gg 1/k$.

If $m\Delta z$ is large enough such that $Y_1$ and $J_1$ are in their
asymptotic
region\footnote{$\sqrt{t}J_1(t)$ is very close to its asymptotic
form $\sqrt{\frac{2}{\pi}}\cos{(t-\frac{3\pi}{4})}$ at $t \ge 3$, while the
small argument behavior $J_1(t) \sim t^2/8$ is a good approximation with
$t \le 1.2$. The first few zeros of $J_1(t)$ are at
$t= 3.83,$ 7.01, 10.17, 13.32, 16.47.}, while $m/k \ll 1$ is still
satisfied, the l-h-s of eqn.~(\ref{quantization})
becomes $\tan{(m\Delta z-\frac{3\pi}{4})}$. Both sides of
eqn.~(\ref{quantization}) can be large provided that:
\beq
\tan{(mz_0)}  \sim - J_1(m/k) / J_2(m/k) \sim -\frac{4k}{m}, 
\eeq
and 
\beq
\tan(m\Delta z-\frac{3\pi}{4}) \gg 1,
\eeq
which implies that 
\beq 
\labell{relativesizes}
\begin{array}{ccc}
mz_0 & \simeq &  \frac{\pi}{2} (2n+1), \\
m\Delta z & \simeq & \frac{\pi}{2} (2l+1) + \frac{3\pi}{4}, 
\end{array}
\eeq 
where the integers $n$ and $l$ can be different, depending on the ratio
between $\Delta z$ and $z_0$. 

The normalization contribution to the wave function from the AdS region
takes the following form for
$m\Delta z$ being large ($\gsim 5$):
\beq
\labell{I1plane}
I_1 \simeq  \frac{N_m^2 L_{m}^2}{\pi} \frac{k}{m}\Delta z. 
\eeq
The normalization contribution from the box gives
\beq
\labell{I2}
I_2  \simeq  B_m^2 z_0 \simeq  \frac{N_m^2 L_{m}^2}{4}\frac{m^2}{k^2} z_0, 
\eeq
where $\tan(mz_0) \sim - 4k/m$ from the quantization condition has been
applied. It is obvious that in the limit $\Delta z \sim z_0 \gg 1/k$, $I_1$
is always much larger than $I_2$ for light modes ($m \ll k$), thus the
AdS behavior dominates the normalization.

One can then calculate the ratio $|\psi_m|^2/|\psi_0|^2$ on the Planck brane, 
\beq 
\rho(m)
\simeq \frac{2\pi}{64}\frac{mz_0}{ k\Delta z} \frac{m^4}{k^4} 
\simeq {\cal O}(\frac{m^5}{k^5})  \;.
\eeq
For these modes there is indeed an enormous suppression at the Planck brane!  

Unfortunately, it is not possible to adjust parameters such that
{\it all} massive modes are suppressed on the Planck brane.
By choosing different $z_0/\Delta z$ as implied by
eqn.~(\ref{relativesizes}), we have changed the periodicity of the Bessel
functions, \ie , the r-h-s of the eqn.~(\ref{quantization}) as a function
of $(m z_0)$ has a periodicity of $\pi z_0/\Delta z$, while the l-h-s is
approximately periodic in $(m z_0)$ with periodicity $\pi$.   As a result,
between two adjacent values $m z_0$ given by eqn.~(\ref{relativesizes}) which
satisfy eqn.~(\ref{quantization}) and yield large $L_m$, the l-h-s and
the r-h-s of eqn.~(\ref{quantization}) meet many times, and inevitably
some of these solutions yield modes with small $L_m$. As we discussed earlier,
when $L_m$ is small, $(L_m \ll (k/m)^4)$, the mode is
unsuppressed at the Planck brane.  

\section{Collider Phenomenology} 

%
The typical cross sections for the total production of on-shell massive
gravitons on the Planck brane is given by \cite{Arkani}:
\beq
\sigma \sim {1\over M_{\rm Planck}^2}\int ^{E}_0 dm\;\rho(m) \quad ,
\eeq
up to dimensionless couplings and numerical factors. 
Here the integral over the density of states extends up
to some maximum kinematically available energy scale $E$.
We assume for simplicity an infinite AdS region, and restrict
$m/k \ll 1$ over the entire integration region.
For a large box, \ie\ $kz_0 \gg 1$, the leading order
contribution to the integral over the density of states
comes from the resonant modes:
\beq
\int ^{E}_0 dm\;\rho(m) = {2\over \pi}Ez_0 + O(E/k)^3 \quad .
\eeq
This result of course mimics an ADD scenario with one
extra dimension having the size of $z_0$.  
Even for a millimeter-size box (the largest allowed by the
Cavendish type bounds), the $z_0$
wave function enhancement of the cross section
cannot overcome the Planck suppression of the couplings.
Thus if we live on the Planck brane there are no observable collider
effects from on-shell production of bulk modes.

One obvious extension of our construction is to change the location
of the Standard Model fields. Here we can imagine confining the Standard
Model to a probe three-brane, as in the scenario of Lykken and
Randall\cite{Lykken}. Locating the probe brane inside the box would not
seem to lead to any new interesting physics since, as seen in
eqn.~\reef{psim},
the suppression (or enhancement) of the bulk modes levels off for $|z|<z_0$.
Hence we consider locating the probe brane at some finite distance $\Delta z_p
=z_p-z_0$ inside the infinite
 AdS region. Here one can profit from the AdS geometry to
generate an interesting hierarchy of scales\cite{small}. As shown
in \cite{Lykken}, the very light continuum bulk modes which contribute
to the Newtonian potential on the brane have wave functions which
simply track the zero-mode. In our case all of the resonant modes
also track the zero-mode, so the complete result for the Newtonian
potential on the probe brane is as given in  eqn.~\reef{totalcorr}.

It was also shown in \cite{Lykken} that the continuum bulk modes
which are heavier than $1/(k{\Delta z_p}^2)$ have wave functions
at the probe brane which are dominated by the $J_2$ rather than $Y_2$
behavior. These modes do not track the zero-mode, and have
large wave function enhancements, leading to potentially
observable collider effects. This behavior still holds in
the case we are considering, precisely for the nonresonant gravity
modes, \ie\ the modes which do not have $L_m \rightarrow 0$
and thus constitute the nonresonant portion of the bulk continuum.
The total cross section for production of these modes on the
probe brane is given by:
\beq
\sigma \sim (2kz_0+1){k^2\over M_{\rm Planck}^2}\, E^6\, 
{\Delta z_p}^8
\labell{crost}
\eeq
where the result is presented for $k{\Delta z_p}\gg1$,
as well as $m/k\ll 1$.
In the limit that $kz_0\rightarrow 0$, this expression 
reduces to precisely that derived in \cite{Lykken}. However,
the factor of $(2kz_0+1)$ in eq.~\reef{crost} provides 
a relative enhancement for $kz_0\gg1$, \ie, for
a large box. The appearance of this factor can be traced
to a suppression of the zero-mode wave function (relative 
to that of the continuum bulk modes) in this scenario,
as seen in eqns.~\reef{psinot} and \reef{normnot}.
The same suppression then effects the definition of $M_{\rm Planck}$ in
eqn.~\reef{plank}.

The net result is that the existence of a large box
can dramatically enhance the collider signals on a probe
brane in the infinite AdS region. The enhancement is not due
to resonant mode production (which as we have seen is
Planck suppressed), but rather to the enhanced production of
the continuum bulk modes. For example, suppose
that $k \sim M_{\rm Planck}$ and that we have a millimeter-sized
box. Then $kz_0$ is a huge enhancement factor:
$kz_0 \simeq 10^{16}$. This would imply that, even with
a probe brane cutoff $1/\Delta z_p$ as large as $10^4$ TeV,
collider effects are suppressed by no more than $E^6/({\rm TeV})^8$.
Hence, in the large box scenario, the production of these bulk modes 
could be within the reach of collider experiments in the forseeable
future. One might also expect that this enhanced production may have
observable astrophysical and cosmological implications.

\section{Discussion}

To summarize, we have presented a brane construction 
which smoothly interpolates between the physics of the ADD and RS
scenarios. Essentially, our five-dimensional background geometry
consists of a slice of flat Minkowski space, ``the box,'' glued between two
AdS regions. The discontinuity in the extrinsic curvature across the
gluing surfaces is interpreted in terms of positive-tension three-branes
located at these positions.
For the most part in the following discussion, we will explicitly comment
on the case of an infinite
fifth dimension, however, most of the comments
carry over to the case where the fifth dimension is finite.

If the size of the box is small compared to the AdS scale,
\ie, $z_0\lsim1/k$, then the low-energy physics is essentially
the same as in the RS scenario. That is, the coupling
of the bulk gravity modes is still essentially controlled by the
AdS scale, and so the corrections to the gravitational potential
\reef{newtan} have precisely the same power law form as in \cite{big}.
Even if $z_0\sim1/k$, there would essentially be only one scale in
the potential \reef{V(z)}, and so one should not find
any radical departures from the RS scenario.

This small box regime, \ie, $z_0\lsim1/k$, would model the situation
of a thick or smooth
Planck brane\cite{thick}. That is a construction where one might attempt
to realize the RS scenario using a smooth domain wall solution to replace
the infinitely thin Planck brane. One would expect (at least naively) that
since, in such a scenario,
the AdS curvature and the thickness of the brane would be determined by
the same underlying microscopic theory, both of these scales would be
of the same order. Our results in the small box regime then agree with
the investigations in \cite{thick}, where it was found that thickening
the Planck brane produced no significant differences from the low-energy
physics of the RS scenario.

In the large box regime, \ie, $z_0\gg1/k$, a new large length scale
is introduced and we do find significant changes in the low energy
physics. In particular, there are resonant modes with enhanced
support inside the box, and so with an enhanced coupling to the
the three-branes. 
Even though there is a continuum of bulk modes with masses
near the resonant mass, their net contribution mimics a single
normalizable mode with this resonant mass. Thus the details of the
AdS region are suppressed in this regime, and to leading order 
in $1/kz_0$ the
brane-boundary at $z=z_0$ simply acts like an orbifold surface. 
That is, the leading order corrections to the Newtonian potential are
identical to those as if we were considering a compactification of flat
five-dimensional spacetime on $S^1/Z_2$.

Recently, H. Verlinde\cite{herman} proposed an interesting way to realize the
RS scenario in superstring theory. This proposal was later elaborated on in
\cite{elab}. Essentially the five-dimensional AdS geometry arises in
the throat geometry near a cluster of D3-branes, while the five-form charge
of the D-branes is absorbed by the ``topology'' of the compactification
geometry in which they are positioned. The Standard Model fields would 
live on a probe brane sitting in the AdS throat, analogous to \cite{Lykken}.
Like our ``gravity-in-a-box'' model, this scenario then has two independent
scales, the AdS curvature scale $1/k$ and the size of the compactification
manifold $L$. Further the latter size must be larger than the AdS scale in
order that the throat of the D3-branes can fit inside this geometry.
Hence we expect that our large box scenario may be closely related to the
low-energy physics of Verlinde's construction.
In particular then, as in
eqn.~\reef{plank} with large $kz_0$, the relation between the observed
four-dimensional Planck scale and the fundamental scale of gravity will be
essentially the same as in the standard ADD scenario. 
This simple relation arises because the normalization of the graviton
zero-mode is dominated
by the integration over the volume of the flat box, external to the AdS
region. As observed in section 7 then, the latter also results in
the production rate of continuum bulk gravitons being enhanced. We expect
this enhancement will be a general feature of the superstring constructions,
and so provide interesting phenomenological constraints for these models

It is interesting to consider the generalization of our brane construction
to spacetimes with more than five dimensions. The RS scenario was generalized
to higher dimensions using intersecting branes in \cite{more}. This discussion
was extended to considering both intersecting branes and different
cosmological
constants in the distinct regions between the branes \cite{mormor}.
Given these results, it is clear that there is no obstacle to extending the
present scenario to higher dimensions. One would have a finite
portion of flat space surrounded by various AdS regions. If the size
of the box is still characterized by a single scale,
we expect that much of the previous discussion would carry over to the
present situation. If the box is smaller than the curvature
scale of the surrounding AdS regions, that the low-energy physics would
be essentially the same as in generalized RS construction of \cite{more}.
If the size of the box is much bigger than the AdS scale, there should
be resonant modes so that the low-energy theory imitates a flat
space ADD scenario. We expect that just as the ADD scenarios are
more phenomenologically interesting in more than five dimensions,
the higher dimensional extensions of our ``gravity-in-a-box'' model
would yield a richer phenomenology. It may be of interest to examine
how to distinguish the low-energy physics of the ADD scenario from
that of a large higher dimensional box. One interesting possibility for our
higher dimensional constructions is that one can engineer a box with
an essentially arbitrary shape in the extra dimensions. In the large
box regime, such a configuration should give rise to a unique spectrum
of masses which would distinguish it from a conventional ADD scenario. \\

{\bf Acknowledgements}

\noindent
Research by JL and JW was supported by the U.S.~Department of Energy
Grant DE-AC02-76CHO3000.
Research by RCM was supported by NSERC of Canada and Fonds FCAR du Qu\'ebec.




\end{document}